\documentclass{ptephy_v1}

\usepackage{amssymb}
\usepackage{amsmath,amscd}
\usepackage{color}
\usepackage{mathrsfs}

\usepackage{bm}

\usepackage[normalem]{ulem}

\begin{document}

\title{Holographic entanglement entropy of deSitter braneworld with Lovelock}


\author{Kouki Kushihara${}^1$}
\author{Keisuke Izumi${}^{2,1}$}
\author{Tetsuya Shiromizu${}^{1,2}$}

\affil{${}^1$Department of Mathematics, Nagoya University, Nagoya 464-8602, Japan}
\affil{${}^2$Kobayashi-Maskawa Institute, Nagoya University, Nagoya 464-8602, Japan}


\begin{abstract}
We examine the deSitter entropy in the braneworld model with the Gauss-Bonnet/Lovelock terms. 
Then, we can see that the deSitter entropy computed through the Euclidean action exactly coincides with the 
holographic entanglement entropy. 
\end{abstract}

\subjectindex{E0}

\maketitle

%
\section{Introduction}

In the development of adS/CFT correspondence \cite{Maldacena, Gubser:1998bc, Witten:1998qj}, 
a remarkable one is the Ryu-Takayanagi proposal for the holographic entanglement entropy \cite{RT2006}. 
This is regarded as a natural extension of the Bekenstein-Hawking entropy for the black hole to general 
cases based on the holographic aspect. 

On the other hand, the braneworld model inspired by string theory also has the holographic feature \cite{bw, adscftbw}. 
Therefore, it is natural to consider the holographic entanglement entropy in the braneworld context too. 
Recent hybrid formulation of adS/CFT(or adS/BCFT \cite{adsbcft}) and braneworld, say Island formula, may be able to 
offer the solution to the information loss paradox in black hole evapolation \cite{island} 
(See also Refs \cite{myers2020a,myers2020b,myers2020c}). 
In this sense, the braneworld setup contributes to understanding the quantum gravity. 

In this paper, we revisit the holographic entanglement entropy of the deSitter braneworld (See Ref. \cite{emparan2006} for black hole). 
In Ref. \cite{Iwashita}, the authors showed the exact agreement between the deSitter entropy computed from the 
Eucliedan path integral \cite{gibbons-hawking} and the Ryu-Takayanagi formula for holographic entanglement entropy 
(See also \cite{Hawking}). However, they also founds a disagreement between them in the braneworld model with the Gauss-Bonnet term. 
This is not surprising result because 
the Ryu-Tayakanagi formula should be improved for the higher derivative theories. Indeed, 
motivated by the formula for the black hole entropy in the Lovelock gravity theory \cite{Jacobson}, 
the authors in Ref. \cite{Myers2011} proposed the new formula which has the correction terms to the Ryu-Takayanagi formula. 
Then, our purpose is to confirm that the formula given in Ref. \cite{Myers2011} coincides with the deSitter entropy in 
braneworld with the Gauss-Bonnet/Lovelock terms. 

The remaining part of this paper is organized as follows. In Sect.~\ref{SecGB}, we describe the braneworld model with the Gauss-Bonnet 
term and deSitter brane in the $n$-dimensional anti-deSitter spacetime. 
We give the detail of the calculation for
 the deSitter entropy through the Euclidean path integral and the holographic entanglement entropy. 
In Sect.~\ref{SecLL}, the analysis and results in the braneworld model with the Lovelock gravity are shown. 
Finally, in Sect.~\ref{Secsum}, we give short summary. 

%
%

\section{Braneworld with Gauss-Bonnet} \label{SecGB}

In this paper, we consider the $Z_2$-symmetric braneworld model in the anti-deSitter bulk. 
The bulk Gauss-Bonnet term \cite{myers1987} is analysed first. 
The system is composed of the $n$-dimensional bulk $(M_n^\pm, g_{MN})$ and the brane $(M_{n-1}, q_{\mu\nu})$ 
(Ref. \cite{Iwashita} focused on the $n=5$ case. Here, we consider any dimensions with $n \geq 5$). The action is 
given by 
\begin{equation}
S=\frac{1}{16\pi G_n}\int_{M_n^+ \cup M_n^-} d^n x{\sqrt {-g}}\Bigl(R-2\Lambda+\frac{\beta \ell^2}{4}{\cal L}_{\rm GB}   \Bigr) + 
\int_{M_{n-1}}d^{n-1}x{\sqrt {-q}}\Bigl(-\sigma +\frac{1}{16\pi G_n}[Q]^- \Bigr),
\end{equation}
where $G_n$ is the $n$-dimensional Newton constant, $R$ is the $n$-dimensional Ricci scalar, $\Lambda$ is a negative cosmlogical constant, 
$\ell$ is supposed to be the anti-deSitter curvature length and $\beta$ is a dimensionless constant. 
Here, the Gauss-Bonnet Lagrangian ${\cal L}_{\rm GB}$ is given as
\begin{equation}
{\cal L}_{\rm GB}=R^2- 4R_{AB}R^{AB}+R_{ABCD}R^{ABCD}
\end{equation}
and the gravitational surface term $Q$ is written in
\begin{equation}
Q=2K+\beta \ell^2 (J-2{}^{(n-1)}G_{\mu\nu}K^{\mu\nu}),
\end{equation}
where ${}^{(n-1)}G_{\mu\nu}$ is the $(n-1)$-dimensional Einstein tensor, $J$ is the trace of $J_{\mu\nu}$ defined by 
\begin{equation}
J_{\mu\nu}=-\frac{1}{3}(2K_{\mu\alpha}K_{\nu\beta}K^{\alpha\beta}-2KK_{\mu\alpha}K_\nu^\alpha
-K_{\mu\nu}K_{\alpha\beta}K^{\alpha\beta}+K^2K_{\mu\nu})
\end{equation}
and $K_{\mu\nu}$ is the extrinsic curvature of $M_{n-1}$ (whose the normal direction is taken to outward for $M_n^+$). 
Supposing that the locus of the brane is $y=0$ in the Gaussian normal coordinate $(y, x^\mu)$ around the brane, 
$[F]^-$ is defined by 
\begin{equation}
[F]^-:=\lim_{y \to +0}F-\lim_{y \to -0}F.
\end{equation}
Then, the bulk field equation is 
\begin{equation}
G_{MN}+\Lambda g_{MN}+\frac{\beta\ell^2}{2}H_{MN}=0,
\end{equation}
where 
\begin{equation}
H_{MN}=RR_{MN}-2R_{MK}R^K_N-2R^{KL}R_{MKNL}+R_{MKLP}R_N^{~KLP}-\frac{1}{4}g_{MN}{\cal L}_{\rm GB}.
\end{equation}
The junction condition is 
\begin{equation}
[K^\mu_\nu-\delta^\mu_\nu K]^-+\frac{\beta \ell^2}{2}[3J^\mu_\nu-\delta^\mu_\nu J-2P^\mu_{~\alpha\nu\beta}K^{\alpha\beta}]^-
=8\pi G_n \tau^\mu_\nu,
\end{equation}
where 
\begin{equation}
P_{\mu\alpha\nu\beta}={}^{(n-1)}R_{\mu\alpha\nu\beta}-2{}^{(n-1)}R_{\mu [\nu}q_{\beta ]\alpha}
+2{}^{(n-1)}R_{\alpha [\nu}q_{\beta ]\mu}+{}^{(n-1)}Rq_{\mu [\nu}q_{\beta ]\alpha}
\end{equation}
and $\tau_{\mu\nu}$ is the energy-momentum tensor on the brane. Since we focus on the vacuum brane, $\tau_{\mu\nu}=-\sigma q_{\mu\nu}$, where 
$\sigma$ is the brane tension. 
 
Hereafter we consider the deSitter brane in the $n$-dimensional anti-deSitter spacetime (${\rm adS}_n$) 
which is the solution to the current model. The bulk metric is given by \cite{Spradlin}
\begin{eqnarray}
ds^2 & = & dr^2+(\ell H)^2 \sinh^2 (r/\ell)[-dt^2+H^{-2}\cosh^2 (Ht)d \Omega_{n-2}^2] \nonumber \\
& = & dr^2+(\ell H)^2 \sinh^2(r/\ell) [-(1-H^2\rho^2)dT^2+(1-H^2\rho^2)^{-1}d\rho^2+\rho^2d \Omega_{n-3}^2 ],\label{dsbw}
\end{eqnarray}
where $\ell$ is the curvature length of ${\rm adS}_n$ and $H$ is the Hubble constant on the brane. Then, 
supposing that the brane is located at $r=r_0$, we see 
\begin{eqnarray}
H^{-1}=\ell \sinh (r_0/\ell).
\end{eqnarray}
The bulk field equation and junction condition imply us 
\begin{eqnarray}
\Lambda=-\frac{(n-1)(n-2) }{2 \ell^2} \left(1 - \beta\frac{(n-3)(n-4)}{4} \right)
\end{eqnarray}
and
\begin{eqnarray}
\frac{(n-2)}{\ell}\cosh (r_0/\ell)\left[1-\beta \frac{(n-3)(n-4)}{6} \left( 1 -\frac{2}{\sinh^2(r_0/\ell)} \right) \right]
=4\pi G_n \sigma,
\end{eqnarray}
respectively. 

%
\subsection{deSitter entropy with Gauss-Bonnet}

In this subsection, we will give the deSitter entropy in the braneworld with the Gauss-Bonnet term. 
One can compute the deSitter entropy through the Euclidean action 
\begin{eqnarray}
I_E & = & \frac{1}{16\pi G_n} \int_{M^+_n \cup M^-_n}d^nx{\sqrt {g}}\Bigl(2\Lambda-R-\frac{\beta \ell^2}{4}{\cal L}_{\rm GB} \Bigr)
+\int_{M_{n-1}}d^{n-1} x{\sqrt {q}}\Bigl(\sigma-\frac{[Q]^-}{16\pi G_n} \Bigr) \nonumber \\
& = &  \frac{(n-1) [2-\beta(n-2)(n-3)]}{ 16 \pi G_n \ell^2}  \int_{M^+_n \cup M^-_n}d^nx{\sqrt {g}} \nonumber \\
&&\hspace{10mm}
-\frac{\coth (r_0/\ell)}{8 \pi G_n \ell} \left[2-\beta(n-2)(n-3) \left( 1-\frac{2}{\sinh^2 (r_0/\ell)} \right)  \right]\int_{M_{n-1}}d^{n-1}x{\sqrt {q}} \nonumber \\
& = & -\frac{(n-2)  \ell^{n-2}}{ 4 \pi G_n } \Omega_{n-1}  \Biggl( 
\left[1-\beta\frac{(n-2)(n-3)}{2}\right] \int_0^{r_0/\ell} dx \sinh^{n-3}x  \nonumber \\
&&\hspace{45mm}
 + \beta (n-3) \sinh^{n-4} (r_0/\ell) \cosh (r_0/\ell)  \Biggr),
\end{eqnarray}
where we used 
\begin{eqnarray}
[Q]^-=  \frac{4(n-1)}{\ell}\coth (r_0/\ell) \left[1-\beta \frac{(n-2)(n-3)}{6} \left(1 -\frac{2}{\sinh^2 (r_0/\ell)} \right) \right].
\end{eqnarray}
and $\Omega_{n-1}$ is the surface area of the $n$ dimensional unit sphere,
\begin{eqnarray}
\Omega_{n-1} = \frac{2 \pi^{n/2}}{\Gamma (n/2)}.
\end{eqnarray}
Then, we see that the deSitter entropy in the braneworld with the Gauss-Bonnet term is given by 
\begin{eqnarray}
S_{\rm dS}
& = & -I_E\nonumber \\ 
& = & \frac{(n-2)  \ell^{n-2}}{ 4 \pi G_n } \Omega_{n-1}  \Biggl(  \left[1-\beta\frac{(n-2)(n-3)}{2}\right] \int_0^{r_0/\ell} dx \sinh^{n-3}x 
\nonumber \\
&&\hspace{45mm}
 + \beta (n-3) \sinh^{n-4} (r_0/\ell) \cosh (r_0/\ell) \Biggr).\label{dSentropy}
\end{eqnarray}

%
\subsection{Holographic entanglement entropy with Gauss-Bonnet}

For the current setup, following Ref. \cite{Jacobson}, we consider
\begin{eqnarray}
S_{\rm JM}=\frac{1}{4G_n}\int_{\Gamma^+ \cup \Gamma^-} d^{n-2}x {\sqrt {h}}\Bigl(1+\frac{\beta \ell^2}{2}{}^{(n-2)}R \Bigr)
+\frac{1}{2G_n}\int_{\partial \Gamma}d^{n-3}x{\sqrt {p}}\frac{\beta\ell^2}{2}[{}^{(n-3)}k]^-, \label{area2}
\end{eqnarray}
where $h_{ij}$ is the induced metric of $(n-2)$-dimensional surface $\Gamma^\pm$ with $\partial \Gamma^+=\partial \Gamma^-=:\partial \Gamma$, 
${}^{(n-2)}R$ is the Ricci scalar of $h_{ij}$, $p_{AB}$ is the induced metric of $\partial \Gamma$ and ${}^{(n-3)}k$ is 
the extrinsic curvature of $\partial \Gamma$. 
We suppose $\Gamma^+ \subset M_n^+$, $\Gamma^- \subset M_n^-$ and $\partial \Gamma \subset M_{n-1}$ 
in the braneworld setup. When $\beta=0$, Eq. (\ref{area2}) becomes to be proportional to the volume of $\Gamma^+ \cup \Gamma^-$. 

Then, one takes the variation of $\Gamma$ for $S_{\rm JM}$ and the minimum value gives us the holographic entanglement entropy. 
From the variation, one obtains 
\begin{eqnarray}
{}^{(n-2)}k-\beta \ell^2 {}^{(n-2)}G_{ij}{}^{(n-2)}k^{ij}=0 \label{variation}
\end{eqnarray}
in the bulk and 
\begin{eqnarray}
\beta \left[{}^{(n-3)}k^{AB} - {}^{(n-3)}k p^{AB} \right]^-  {}^{(n-2)}k_{AB} =0 \label{variation2}
\end{eqnarray}
on the brane, 
where ${}^{(n-2)}k_{ij}$ and ${}^{(n-3)}k_{AB}$ are the extrinsic curvatures of $\Gamma$ and of $\partial \Gamma$ in $\Gamma$. 
These equations determine the geometry of $\Gamma$. When $\beta=0$, Eq. (\ref{variation}) becomes ${}^{(n-2)}k=0$ and 
Eq. (\ref{variation2}) becomes trivial. 

For the deSitter braneworld, it is easy to see that the 3-surface $\Gamma_\ast$ with $T=$const. and $\rho=H^{-1}$ in Eq. (\ref{dsbw}) 
satisfies Eq. (\ref{variation}). This is because $\Gamma_\ast$ is the hyperboloid with the curvature length $\ell$, that is, 
the induced metric is $h=dr^2+\ell^2 \sinh^2(r/\ell)d\Omega_{n-3}^2$, and then 
Eq. (\ref{dsbw}) implies $[1-\beta(n-3)(n-4)/2]{}^{(n-2)}k$=0. In Ref. \cite{Iwashita}, 
it was shown that $\Gamma_\ast$ satisfies  Eq. (\ref{variation}). 
Moreover, since ${}^{(n-2)}k_{ij } =0$ is satisfied on $\Gamma_\ast$, we also see that Eq. (\ref{variation2}) is trivially satisfied. 

Since ${}^{(n-2)}R=-(n-2)(n-3)/\ell^2$, ${}^{(n-3)}k=[(n-3)/\ell]\coth (r_0/\ell)$, the holographic entanglement entropy is computed as 
\begin{eqnarray}
S_{\rm JM}& = & \frac{1}{4G_n} \left(1-\beta\frac{(n-2)(n-3)}{2} \right) \int_{\Gamma_\ast}d^{n-2}x{\sqrt {h}}
\nonumber \\
&& \hspace{10mm}
+\frac{\beta \ell (n-3)}{2 G_n}\coth(r_0/\ell) \int_{\partial \Gamma_\ast} d^{n-3}x{\sqrt {p}}
\nonumber \\
& = & \frac{\ell^{n-2}}{2G_n} \Omega_{n-3} \Biggl( 
\left[1-\beta\frac{(n-2)(n-3)}{2} \right]  \int_0^{r_0/\ell} dr  \sinh^{n-3}x 
\nonumber \\
&& \hspace{30mm}
+\beta (n-3)  \sinh^{n-4}(r_0/\ell) \cosh(r_0/\ell)
 \Biggr) , \label{EE}
\end{eqnarray}
where $\Omega_{n-3}$ is the surface area of the $(n-2)$ dimensional unit sphere. 
Since $\Omega_{n-3}$ is expressed  with $\Omega_{n-1}$,
\begin{eqnarray}
\Omega_{n-3} = \Omega_{n-1} \frac{(n-2)}{2\pi} \label{Omegatrans}
\end{eqnarray}
we can find that 
\begin{eqnarray}
S_{\rm dS}=S_{\rm JM}
\end{eqnarray}
holds exactly!

%
%

\section{Braneworld with Lovelock}\label{SecLL}

We will consider more generic setting, the braneworld model with the Lovelock terms. 
The analysis can be proceeded in the same way as in the case with the Gauss-Bonnet term. 
We follow the definitions of geometrical quantities and objects given in the previous section. 
The action with boundary is given in Ref.~\cite{myers1987}, 
\begin{eqnarray}
S & = & \frac{1}{16\pi G_n} \int_{M^+_n \cup M^-_n}d^nx {\sqrt {-g}}\Bigl(-2\Lambda + \sum_m c_m {\cal L}_{m} \Bigr)
\nonumber \\
&&\hspace{40mm} +\int_{M_{n-1}}d^{n-1} x{\sqrt {-q}}\Bigl(-\sigma+ \sum_m c_m \frac{[Q_m]^-}{16\pi G_n} \Bigr),
\label{LLaction}
\end{eqnarray}
where $c_m$'s are coefficients of Lovelock terms,
\begin{eqnarray}
{\cal L}_{m}
& = &\frac{ 1}{2^m}  g^{K_1L_1\ldots K_m L_m}_{M_1 N_1 \ldots M_m N_m} R_{K_1L_1}{}^{M_1N_1} \ldots R_{K_mL_m}{}^{M_m N_m}, \\
Q_m
& = & \frac{4 m}{2^m}  \int_0^1 ds\, q^{\alpha_1 \beta_1 \ldots \alpha_{m-1} \beta_{m-1} \alpha_m}_{\mu_1 \nu_1\ldots \mu_{m-1} \nu_{m-1}\mu_m}  
\left( {}^{(n-1)}R_{\alpha_1\beta_1}{}^{\mu_1\nu_1} -2s^2 K^{\mu_1}_{\alpha_1} K^{\nu_1}_{\beta_1} \right) 
\nonumber \\
&&\hspace{40mm}
 \ldots  \left({}^{(n-1)} R_{\alpha_{m-1}\beta_{m-1}}{}^{\mu_{m-1} \nu_{m-1}}  -2s^2 K^{\mu_{m-1}}_{\alpha_{m-1}} K^{\nu_{m-1}}_{\beta_{m-1}} \right) 
K^{\mu_m}_{\alpha_m}
\nonumber \\
& = & \frac{ 4m}{2^m}   \, q^{\alpha_1 \beta_1 \ldots \alpha_{m-1} \beta_{m-1} \alpha_m}_{\mu_1 \nu_1\ldots \mu_{m-1} \nu_{m-1}\mu_m}   
\sum_{k=0}^{m-1} \binom{m-1}{k} \frac{(-2)^k}{2k+1} K^{\mu_1}_{\alpha_1} K^{\nu_1}_{\beta_1} \ldots K^{\mu_k}_{\alpha_k} K^{\nu_k}_{\beta_k}K^{\mu_m}_{\alpha_m}
\nonumber \\
&&\hspace{45mm}
  {}^{(n-1)} R_{\alpha_{k+1}\beta_{k+1}}{}^{\mu_{k+1}\nu_{k+1}} \ldots {}^{(n-1)}  R_{\alpha_{m-1}\beta_{m-1}}{}^{\mu_{m-1} \nu_{m-1}} ,
\nonumber \\\end{eqnarray}
and $\binom{m-1}{k}$ is the binomial coefficients. 
The tensors $ g^{K_1L_1\ldots K_m L_m}_{M_1 N_1 \ldots M_m N_m}$ and $q^{\alpha_1 \beta_1 \ldots \alpha_{m-1} \beta_{m-1} \alpha_m}_{\mu_1 \nu_1\ldots \mu_{m-1} \nu_{m-1}\mu_m}  $ are defined as
\begin{eqnarray}
&&g^{K_1L_1\ldots K_m L_m}_{M_1 N_1 \ldots M_m N_m}
 := 
(2m)! \,  \delta^{K_1}_{[M_1} \delta^{L_1}_{N_1}\ldots \delta^{K_m}_{M_m} \delta^{L_m}_{N_m]} ,\\
&&q^{\alpha_1 \beta_1 \ldots \alpha_{m-1} \beta_{m-1} \alpha_m}_{\mu_1 \nu_1\ldots \mu_{m-1} \nu_{m-1}\mu_m} 
:= 
(2m-1)! \,  q^{\alpha_1}_{[\mu_1} q^{\beta_1}_{\nu_1}\ldots q^{\alpha_{m-1}}_{\mu_{m-1}} q^{\beta_{m-1}}_{\nu_{m-1}} q^{\alpha_m}_{\mu_m]} .
\end{eqnarray}
The brackets $[M_1 \ldots N_m]$ in index indicate antisymmetrization, for instance, $T_{[MN]} = (1/2!) (T_{MN} - T_{NM}) $.
The sum in Eq. (\ref{LLaction}) is taken from $m=0$ to $[n/2]$, where $[...]$ is the floor function. 

The equation of motion and the junction condition are obtained by taking the variation of Eq. (\ref{LLaction}).
The equation of motion becomes 
\begin{eqnarray}
\Lambda g_{MN} +\sum_m c_m E^{(m)}_{MN}=0,
\end{eqnarray}
where 
\begin{eqnarray}
E^{(m)}_{MN} = \frac{ m}{2^m} \, g^{K_1L_1\ldots K_m L_m}_{M_1 N_1 \ldots M_m N_m} g_{K_1 M} R_{N L_1}{}^{M_1N_1} 
R_{K_2L_2}{}^{M_2 N_2} \ldots R_{K_mL_m}{}^{M_m N_m}
-\frac12 {\cal L}_{m} g_{MN} .
\end{eqnarray}
This gives us 
\begin{eqnarray}
\Lambda = \frac{1}{2n} \sum_m c_m (n-2m) {\cal L}_{m}. \label{L-LL}
\end{eqnarray}
The junction condition is 
\begin{eqnarray}
16 \pi G_n \sigma q_{\mu\nu} + \sum_m c_m [J^{(m)}_{\mu\nu} ]^- =0,
\end{eqnarray}
where 
\begin{eqnarray}
&&J^{(m)}_{\mu\nu} = \frac{4m}{2^m}
q^{\alpha_1 \beta_1 \ldots \alpha_{m-1} \beta_{m-1} \alpha_m}_{\mu_1 \nu_1\ldots \mu_{m-1} \nu_{m-1}\mu_m}   
\sum_{k=0}^{m-1} \binom{m-1}{k} \frac{(-2)^k}{2k+1} 
\nonumber\\ 
&&\hspace{5mm}
\times \Biggl( (2k+1)
K^{\mu_1}_{\alpha_1} K^{\nu_1}_{\beta_1} \ldots K^{\mu_k}_{\alpha_k} K^{\nu_k}_{\beta_k}K^{\mu_m}_{\mu} q_{\alpha_m \nu}
\nonumber\\ 
&&\hspace{40mm}
\times {}^{(n-1)} R_{\alpha_{k+1}\beta_{k+1}}{}^{\mu_{k+1}\nu_{k+1}} \ldots {}^{(n-1)}  R_{\alpha_{m-1}\beta_{m-1}}{}^{\mu_{m-1} \nu_{m-1}}
\nonumber \\
&&\hspace{10mm}
 +
2 (m-k-1)
  K^{\mu_1}_{\alpha_1} K^{\nu_1}_{\beta_1} \ldots K^{\mu_k}_{\alpha_k} K^{\nu_k}_{\beta_k}K^{\mu_m}_{\alpha_m}
\nonumber \\
&&\hspace{10mm}
\times {}^{(n-1)} R_{\alpha_{k+1}\beta_{k+1}}{}^{\mu_{k+1}\nu_{k+1}} \ldots {}^{(n-1)}  R_{\alpha_{m-2}\beta_{m-2}}{}^{\mu_{m-2} \nu_{m-2}}
 {}^{(n-1)} R_{\alpha_{m-1}\beta_{m-1}}{}^{\mu_{m-1}}{}_\mu\,  q_{\nu}^{\nu_{m-1}} 
 \Biggr)
\nonumber\\ 
&&\hspace{15mm}
- Q_m q_{\mu\nu}.
\end{eqnarray}
Then, the brane tension is expressed as 
\begin{eqnarray}
16 \pi G_n \sigma = \sum_m c_m \frac{n-2m}{n-1} [Q_m]^-. \label{J-LL}
\end{eqnarray}

%
\subsection{deSitter entropy with Lovelock}

Let us calculate the deSitter entropy through the Euclidean action. 
Substituting Eqs. (\ref{L-LL}) and (\ref{J-LL}) to the Euclidean action, we have 
\begin{eqnarray}
I_E & = & \frac{1}{16\pi G_n} \int_{M^+_n \cup M^-_n}d^nx{\sqrt {g}}\Bigl(2\Lambda- \sum_m c_m {\cal L}_{m} \Bigr)
+\int_{M_{n-1}}d^{n-1} x{\sqrt {q}}\Bigl(\sigma- \sum_m c_m \frac{[Q_m]^-}{16\pi G_n} \Bigr) \nonumber \\
& = & -\frac{1}{8\pi G_n} \int_{M^+_n \cup M^-_n}d^nx{\sqrt {g}} \sum_m c_m \frac{m}{n} {\cal L}_{m}
- \frac{1}{16\pi G_n} \int_{M_{n-1}}d^{n-1} x{\sqrt {q}} \sum_m c_m \frac{2m-1}{n-1} [Q_m]^- .
\nonumber \\
\end{eqnarray}
Since the Riemann curvature of bulk is 
\begin{eqnarray}
R_{KLMN} = -\ell^{-2} (g_{KM}g_{LN}-g_{KN}g_{LM}),
\end{eqnarray}
${\cal L}_{m}$ is calculated as 
\begin{eqnarray}
{\cal L}_{m} = (-1)^m \frac{n!}{(n-2m)!} \ell^{-2m}. \label{Lm}
\end{eqnarray}
On the other hand, the Riemann curvature and the extrinsic curvature of the brane are 
\begin{eqnarray}
&&{}^{(n-1)}R_{\alpha\beta\mu\nu} = \ell^{-2} \sinh^{-2} (r_0/\ell) (q_{\alpha\mu}q_{\beta\nu}-q_{\alpha\nu}q_{\beta\mu})  , \\
&&K_{\mu\nu}= \ell^{-1} \coth(r_0/\ell) q_{\mu\nu}
\end{eqnarray}
Then, we have
\begin{eqnarray}
Q_{m} = 2m \frac{(n-1)!}{(n-2m)!} \ell^{-2m+1} \frac{\cosh(r_0/\ell)}{\sinh^{2m-1}(r_0/\ell)} 
\int_0^1 ds \left(1- s^2 \cosh ^2(r_0/\ell) \right)^{m-1}. \label{Qm}
\end{eqnarray}

With Eqs. (\ref{Lm}) and (\ref{Qm}), the deSitter entropy $S_{\rm dS}=-I_E$ is calculated as 
\begin{eqnarray}
S_{\rm dS} 
%
& = &  \frac{\Omega_{n-1}}{4\pi G_n} \sum_m m c_m \ell^{n-2m} \frac{(n-2)!}{(n-2m)!}   \Biggl( (-1)^m 
 (n-1) \int_0^{r_0/\ell} dx \sinh^{n-1} x   
\nonumber \\
&& \hspace{2mm}
+  (2m-1) \cosh(r_0/\ell) \sinh^{n-2m} (r_0/\ell) 
\int_0^1 ds \left(1- s^2 \cosh ^2(r_0/\ell) \right)^{m-1} \Biggr). 
\end{eqnarray}

%
\subsection{Holographic entanglement entropy with Lovelock}

The holographic entanglement entropy is also given in Ref.~\cite{Jacobson}. Then, as subSect. 2.2, we first consider 
\begin{eqnarray}
S_{\rm JM}= \frac{1}{4G_n}\int_{\Gamma^+ \cup \Gamma^-} d^{n-2}x {\sqrt {h}}
\sum_m c_m \tilde {\cal L}_{m} 
+\frac{1}{4G_n}\int_{\partial \Gamma}d^{n-3}x
{\sqrt {p}} \sum_m c_m [ \tilde Q_m]^-, 
\end{eqnarray}
where
\begin{eqnarray}
\tilde {\cal L}_{m}
& = &\frac{ m }{2^{m-1}}  h^{i_1j_1\ldots i_{m-1} j_{m-1}}_{k_1 l_1 \ldots k_{m-1} l_{m-1}} {}^{(n-2)}R_{i_1j_1}{}^{k_1l_1} \ldots {}^{(n-2)} R_{i_{m-1} j_{m-1}}{}^{k_{m-1} l_{m-1}}, \\
\tilde Q_m
& = & \frac{4 m(m-1)}{2^{m-1}}  \int_0^1 ds\, p^{A_1 B_1 \ldots A_{m-2} B_{m-2} A_{m-1}}_{C_1 D_1\ldots C_{m-2} D_{m-2}C_{m-1}}  
\left( {}^{(n-3)}R_{A_1B_1}{}^{C_1D_1} -2s^2 {}^{(n-3)}k^{C_1}_{A_1} {}^{(n-3)}k^{D_1}_{B_1} \right) 
\nonumber \\
&&\hspace{8mm}
\ldots \left({}^{(n-2)} R_{A_{m-2} B_{m-2}}{}^{C_{m-2} D_{m-2}}  -2s^2 {}^{(n-3)}k^{C_{m-2}}_{A_{m-2}} {}^{(n-3)}k^{D_{m-2}}_{B_{m-2}} \right) 
{}^{(n-3)}k^{C_{m-1}}_{A_{m-1}}.
\end{eqnarray}
Here, 
 $  h^{i_1j_1\ldots i_{m-1} j_{m-1}}_{k_1 l_1 \ldots k_{m-1} l_{m-1}}$ and $p^{A_1 B_1 \ldots A_{m-2} B_{m-2} A_{m-1}}_{C_1 D_1\ldots C_{m-2} D_{m-2}C_{m-1}}  $ are defined by
\begin{eqnarray}
&& h^{i_1j_1\ldots i_{m-1} j_{m-1}}_{k_1 l_1 \ldots k_{m-1} l_{m-1}}
 := 
(2m-2)! \,  h^{i_1}_{[k_1} h^{j_1}_{l_1}\ldots h^{i_{m-1}}_{k_{m-1}} h^{j_{m-1}}_{l_{m-1}]} ,\\
&&p^{A_1 B_1 \ldots A_{m-2} B_{m-2} A_{m-1}}_{C_1 D_1\ldots C_{m-2} D_{m-2}C_{m-1}} 
:= 
(2m-3)! \,  p^{A_1}_{[C_1} p^{B_1}_{D_1}\ldots p^{A_{m-2}}_{C_{m-2}} q^{B_{m-2}}_{D_{m-2}} q^{A_{m-1}}_{C_{m-1}]} .
\end{eqnarray}

Then the minimum value of $S_{\rm JM}$ for the variation of $\Gamma$ gives us the holographic entanglement entropy which is evaluated on a surface satisfying
\begin{eqnarray}
\sum_m c_m \tilde E^{(m)}_{ij} {}^{(n-2)}k^{ij} =0 \label{variation3}
\end{eqnarray}
in the bulk and
\begin{eqnarray}
\sum_m c_m  [\tilde J^{(m)}_{AB} ]^- {}^{(n-2)}k^{AB}  =0 \label{variation4}
\end{eqnarray}
on the brane, 
where 
\begin{eqnarray}
&&\!\!\!
\tilde E^{(m)}_{ij} = \frac{ m}{2^{m-1}} \, h^{i_1j_1\ldots i_{m-1} j_{m-1}}_{k_1 l_1 \ldots k_{m-1} l_{m-1}} g_{i_1 i} R_{j j_1}{}^{k_1l_1} R_{i_2 j_2}{}^{k_2 l_2}\ldots R_{i_{m-1} j_{m-1}}{}^{k_{m-1} l_{m-1}}
-\frac12 \tilde {\cal L}_{m} h_{ij} ,\\
&&\!\!\!
\tilde J^{(m)}_{AB} = \frac{4m(m-1)}{2^{m-1}}
p^{A_1 B_1 \ldots A_{m-2} B_{m-2} A_{m-1}}_{C_1 D_1\ldots C_{m-2} D_{m-2} C_{m-1}}   
\sum_{l=0}^{m-2} \binom{m-2}{l} \frac{(-2)^l}{2l+1} 
\nonumber\\ 
&&\hspace{5mm}
\times \Biggl( (2l+1)
 {}^{(n-3)} k^{C_1}_{A_1}  {}^{(n-3)} k^{D_1}_{B_1} \ldots  {}^{(n-3)} k^{C_l}_{A_l } {}^{(n-3)} k^{D_l}_{B_l}   {}^{(n-3)} k^{C_{m-1}}_{A} p_{A_{m-1} B}
\nonumber\\ 
&&\hspace{40mm}
\times {}^{(n-3)} R_{A_{l+1}B_{l+1}}{}^{C_{l+1}D_{l+1}} \ldots {}^{(n-3)}  R_{A_{m-2}B_{m-2}}{}^{C_{m-2} D_{m-2}}
\nonumber \\
&&\hspace{8mm}
 +
2 (m-l-2)
  k^{C_1}_{A_1} k^{D_1}_{B_1} \ldots k^{C_l}_{A_l} k^{D_l}_{B_l} k^{C_{m-1}}_{A_{m-1}}
\nonumber \\
&&\hspace{8mm}
\times {}^{(n-3)} R_{A_{l+1}B_{l+1}}{}^{C_{l+1}D_{l+1}} \ldots {}^{(n-3)}  R_{A_{m-3}B_{m-3}}{}^{C_{m-3} D_{m-3}}
 {}^{(n-3)} R_{A_{m-2}B_{m-2}}{}^{C_{m-2}}{}_{A}\, p_{B}^{D_{m-2}} 
 \Biggr)
\nonumber\\ 
&&\hspace{15mm}
- \tilde Q_m p_{AB}.
\end{eqnarray}
As is the case in Sect.~\ref{SecGB}, 
 on $T=$const. hypersurface, $\rho = H^{-1}$, which is a minimal surface ${}^{(n-2)}k =0$, satisfies Eqs. (\ref{variation3}) and 
(\ref{variation4}). On this surface, we have
\begin{eqnarray}
\tilde {\cal L}_{m}
& = & m \frac{(n-2)!}{(n-2m)!} (-1)^{m-1} \ell^{-2m+2}, \\
\tilde Q_m
& = & 2 m(m-1)  \frac{(n-3)!}{(n-2m)!} \ell^{-2m+3} \sinh^{-2m+3}(r_0/\ell) \cosh(r_0/\ell)
\nonumber \\
&&\hspace{30mm}
\times \int_0^1 ds \bigl(1-s^2 \cosh^2 (r_0/\ell)\bigr)^{m-2}. 
\end{eqnarray}
Then, the holographic entanglement entropy is calculated in 
\begin{eqnarray}
S_{\rm JM}
&=&
\frac{\Omega_{n-3}}{2G_n}   
\sum_m m c_m   \frac{(n-3)!}{(n-2m)!} \ell^{n-2m} \Biggl((-1)^{m-1} (n-2) \int_0^{r_0/\ell} dx \, \sinh^{n-3}x 
\nonumber \\
&&
+2(m-1) \sinh^{n-2m}(r_0/\ell) \cosh(r_0/\ell)
\int_0^1 ds \bigl(1-s^2 \cosh^2 (r_0/\ell)\bigr)^{m-2} \Biggr) 
\nonumber \\
& = & \frac{\Omega_{n-1}}{4\pi G_n} \sum_m m c_m \ell^{n-2m} \frac{(n-2)!}{(n-2m)!}   \Biggl( (-1)^m 
 (n-1) \int_0^{r_0/\ell} dx \sinh^{n-1} x   
\nonumber \\
&& \hspace{2mm}
+  (2m-1) \cosh(r_0/\ell) \sinh^{n-2m} (r_0/\ell) 
\int_0^1 ds \left(1- s^2 \cosh ^2(r_0/\ell) \right)^{m-1} \Biggr) . 
\nonumber \\
\end{eqnarray}
where, in the second equality, we used Eq. (\ref{Omegatrans}) and 
\begin{eqnarray}
&&(n-2) \int_0^{r_0/\ell} dx\, \sinh^{n-3} x
\nonumber \\ 
&& \hspace{10mm}
 = -(n-1) \int_0^{r_0/\ell} dx\, \sinh^{n-1} x +
\sinh^{n-2} (r_0/\ell) \cosh  (r_0/\ell), \\
&&2(m-1) \int_0^1 ds \bigl( 1- s^2 \cosh^2 (r_0/\ell)\bigr)^{m-2}
\nonumber \\ 
&& \hspace{10mm}
=(-1)^m \sinh^{2m-2}(r_0/\ell) + (2m-1) \int_0^1 ds \bigl( 1- s^2 \cosh^2 (r_0/\ell)\bigr)^{m-1}.
\end{eqnarray}

As a summary, we can see that 
\begin{eqnarray}
S_{\rm JM} = S_{\rm dS} 
\end{eqnarray}
holds exactly in the braneworld with Lovelock terms.

%
\section{Summary}\label{Secsum}

In this paper, we revisited the comparison between the holographic entanglement entropy and deSitter entropy 
in the braneworld model with higher-curvature corrections, that is, the Gauss-Bonnet/Lovelock terms. 
Employing the Jacobson-Myers formula for the 
holographic entanglement entropy, we could show the exact agreement of both. These results may encourage to have 
the general formulation for holographic entanglement entropy in the braneworld context.

\ack

T. S. and K. I.  are supported by Grant-Aid for Scientific Research from Ministry of Education, 
Science, Sports and Culture of Japan (No. 17H01091). 
K.\,I. is also supported by JSPS Grants-in-Aid for Scientific Research (B) (20H01902).
We thank Tadashi Takayanagi for useful discussion.


%
%


\end{document}